\documentclass[pra, 12 pt]{revtex4}
\usepackage[english]{babel}
\usepackage{amsmath}
\usepackage{epsfig}

\begin{document}
%\draft
\title{Anomalous Transport in Velocity Space, from Fokker-Planck to General Equation}
\author{S.A. Trigger}
\address{Joint\, Institute\, for\, High\, Temperatures, Russian\, Academy\,
of\, Sciences, 13/19, Izhorskaia Str., Moscow\, 125412, Russia;\\
Humboldt University, Physics Institute, Newtonstr. 15, D-12489 Berlin-Adlershof,
Germany\\email:\,satron@mail.ru}

\begin{abstract}

The problem of anomalous diffusion in momentum (velocity)
space is considered based on the master equation and the
appropriate probability transition function (PTF). The approach recently developed
for coordinate space in [1], is applied with necessary modifications to velocity space.
A new general equation for the time evolution of the momentum distribution
function in momentum space is derived. This
allows the solution of various problems of anomalous
transport when the probability transition function (PTF) has a
long tail in momentum space.

For the opposite cases of the PTF rapidly decreasing
as a function of transfer momenta (when large
transfer momenta are strongly suppressed), the developed
approach allows us to consider strongly non-equilibrium cases
of the system evolution. The stationary and non-stationary
solutions are studied.

As an example, the particular case of the Boltzmann-type PT-
function for collisions of heavy and light
particles with the determined (prescribed) distribution
function, which can be strongly non-equilibrium, is considered within
the proposed general approach. The appropriate
diffusion and friction coefficients are found. The Einstein
relation between the friction and diffusion coefficients is shown to be violated in these cases.
\\

PACS number(s): 52.27.Lw, 52.20.Hv, 05.40.Fb

\end{abstract}

\maketitle

\section{Introduction}

Interest in the anomalous diffusion is prompted by a large variety
of applications, i.e., semiconductors, polymers, some granular systems,
plasmas under specific conditions, various objects in biological
systems, physical-chemical systems, and others.

Deviation from the linear in time dependence $<r^2(t)>\sim t$
of the mean-squared displacement has been experimentally observed,
in particular, under essentially non-equilibrium conditions and
in some disordered systems. The mean-squared separation of a pair
of particles passively moving in a turbulent flow anomalously increases (according
to the Richardson's law) with the third power of time [2]. For
diffusion typical of glasses and related complex systems [3], the
observed anomalous time dependence is slower than linear. These two types of
anomalous diffusion are referred to as superdiffusion,
$<r^2(t)>\sim t^\alpha$ $(\alpha>1)$, and subdiffusion $(\alpha<1)$,
[4]. To describe these two diffusion regimes, a number of
efficient models and methods have been proposed. The continuous
time random walk (CTRW) model of Scher and Montroll [5], leading
to strongly subdiffusive behavior, provides a basis for
understanding photoconductivity in strongly disordered and glassy
semiconductors. The Levy-flight model [6], leading to
superdiffusion, describes various phenomena such as
self-diffusion in micelle systems [7], reaction and transport in
polymer systems [8], and has been even to the stochastic
description of financial market indices [9]. For both cases, the
so-called fractional differential equations in coordinate and time
spaces were successfully applied [10].

However, recently a more general approach has been proposed in
[1,11], which reproduces the results of the standard fractional
differentiation method (when it is applicable) and
allows the describe of more complicated cases of anomalous
diffusion. In [12], this approach has also been applied
to diffusion in a time-dependent external field.

In this paper, the problem of anomalous diffusion in momentum
(velocity) space is considered. In spite of the formal
similarity, diffusion in momentum space is very different
physically from coordinate space diffusion. This is obvious because momentum conservation
which takes place in momentum space has no analogy in coordinate space.

Some aspects of anomalous diffusion in velocity space have
been considered in several papers [13-18]. In general, in comparison with anomalous diffusion in coordinate
space, anomalous diffusion in velocity space is poorly
studied. To our knowledge, there still is no corresponding way to describe
anomalous diffusion in velocity space self-consistently .

In this paper, a new kinetic equation for anomalous diffusion in
velocity space is derived (see also [19,20]) based on the
appropriate expansion of the PTF (in spirit of the approach
proposed in [1] for the diffusion in coordinate space) and some
particular problems are studied on this basis.

The paper is organized as follows. The diffusion in velocity space
for the cases of the normal and anomalous behavior of the PT-function
is presented in Section II. Starting from argumentation
based on the Boltzmann-type of the PTF, we derive a new kinetic
equation which then can be applied not only to the
Boltzmann-type processes, but also to a wide class of processes
occurring in nature with other types of PT-functions. The particular cases
of anomalous diffusion for hard spheres collisions with the
specific power-type prescribed distribution function of light
particles are analyzed in Section III. More general
examples of anomalous diffusion are also considered. The
non-stationary solution for the distribution function
identical at $t=0$ to the initial distribution is found. In
Section IV, the Boltzmann-type equation is used to consider
the effect light particle drift on the PT-function and
on the kinetic equation structure. In Section V, a
short review of anomalous diffusion in coordinate space is
presented. The method and results of this paper can be easily
extended to describe plasma-like systems and anomalous processes in energy space.

\section{Diffusion in velocity space on the basis of
the master-type equation}

Let us now consider the main problem formulated in the
introduction, namely, the description of diffusion in velocity space ($V$-space). Such a description
would be based on a corresponding master equation describing the
balance of grains coming at and from the point $p$ at the instant
$t$. The structure of this equation
\begin{equation}
\frac{df_g({\bf p},t)}{dt} = \int d{\bf q} \left\{W ({\bf q, p+q})
f_g({\bf p+q}, t) - W ({\bf q, p}) f_g({\bf p},t) \right\}.
\label{DC2b}
\end{equation}
is formally similar to the master equation Eq.~(\ref{DC2}) (see
below) in coordinate space. Surely, for coordinate space,
there is no conservation law similar to that in momentum
space. The probability transition function $W({\bf q, p'})$ (the
transition rate) describes the probability that a grain with
momentum ${\bf p'}$ (point ${\bf p'}$) passes from this point
${\bf p'}$ to the point ${\bf p}$ per unit time by transferring
the momentum ${\bf q=p'- p}$ to the surrounding medium. Assuming
initially that the characteristic transfer momentum ${\bf q}$ is smaller than
${\bf p}$, one may expand Eq.~(\ref{DC2b}) in
on ${\bf q}$ to the second order. Then, we arrive at the usual form of the Fokker-Planck equation for the density distribution function $f_g({\bf p},t)$ (see, e.g., [21])

\begin{equation}
\frac{df_g({\bf p},t)}{dt} = \frac {\partial}{\partial p_\alpha}
\left[ A_\alpha ({\bf p}) f_g({\bf p},t) + \frac{\partial}
{\partial p_\beta} \left(B_{\alpha\beta}({\bf p}) f_g({\bf p},t)
\right)\right]. \label{DC3b}
\end{equation}
\begin{equation}
A_\alpha({\bf p}) = \int d^s q q_\alpha W({\bf q, p});\;\;\;\
B_{\alpha\beta}({\bf p})= \frac{1}{2}\int d^s q q_\alpha q_\beta
W({\bf q, p}). \label{DC4b}
\end{equation}
The coefficients $A_\alpha$ and $B_{\alpha \beta}$ describe the
friction force and diffusion, respectively. The indices $\alpha$ and
$\beta$ correspond to the coordinate axes $x, y, z$ (for
dimension $s=3$).

Since the velocity of heavy particles is low, the $\bf
p$-dependence of the PTF can be neglected for calculating the
diffusion term, which in this case is a constant
$B_{\alpha\beta}=\delta_{\alpha\beta}B$. The constant $B$ is equal to the integral
\begin{equation}
B = \frac{1}{2s}\int d^s q q^2 W(q), \label{DC6b}
\end{equation}
where $s$ is the dimension of velocity space ($V$-space0. Neglecting the $\bf
p$-dependence of the PT-function, we arrive at the coefficient
$A_\alpha=0$ (while the diffusion coefficient is constant).

As is well known this neglect is wrong and the coefficient $A_\alpha$
for the Fokker-Planck equation can be determined using the assumption that
argument that the stationary distribution function is Maxwellian.
Then, we arrive at the standard relation between the coefficients $MT
A_\alpha(p)=p_\alpha B$ (here $M$ is the particle mass
and $T$ is temperature of the particles in the equilibrium
state). This relation is an analogue of the Einstein
relation in coordinate space.  However, this
argumentation is not applicable to systems far from equilibrium,.

To find the coefficients in the kinetic equation, which are
appropriate for more general non-equilibrium situations, e.g., for slowly
decreasing PT-functions or (and) for strongly non-equilibrium
kernels $W$, let us use a more general way based on a certain small
parameter (in the simplest case, e.g., on the difference of
velocities of light and heavy particles for Boltzmann-type
collisions). To calculate the function $A_\alpha$ we have
take into account that the function $W(\bf {q,p})$ is a scalar and can
depend only on variables $q, {\bf q \cdot p}, p$. Expanding $W(\bf {q,p})$ in the product
$\bf {q\cdot p}$, one arrives at the approximate representation of
the functions $W(\bf {q,p})$ and $W({\bf q, p+q})$,
\begin{eqnarray}
W({\bf q,p)}\simeq W(q)+\tilde W'(q)({\bf q \cdot p})+
 \frac{1}{2}\tilde W''(q) ({\bf q \cdot p})^2 . \label{DC7b}
\end{eqnarray}
\begin{equation}
W ({\bf q, p+q})\simeq W(q)+\tilde W'(q) \,({\bf q \cdot
p})+\frac{1}{2}\tilde W''(q) ({\bf q \cdot p})^2 +q^2\tilde
W'(q),\label{DC9b}
\end{equation}
Here we introduced the derivatives $\tilde W'(q)\equiv \partial W (q, {\bf q \cdot p}, p)
/\partial ({\bf q p})\mid_{{\bf q \cdot p}=0, \,p=0}$ and $\tilde
W''(q)\equiv \partial^2 W (q, {\bf q \cdot p},p) /\partial ({\bf q
p})^2 \mid_{{\bf q \cdot p}=0,\, p=0}$.

Then, with necessary accuracy (which corresponds to the derivation of the well known Fokker-Planck equation) for the coefficient $A_\alpha$ we find
\begin{equation}
A_\alpha({\bf p}) = \int d^s q q_\alpha q_\beta p_\beta \tilde
W'(q)= p_\alpha \int d^s q q_\alpha q_\alpha  \tilde
W'(q)=\frac{p_\alpha}{s} \int d^s q q^2  \tilde W'(q)\label{DC10b}
\end{equation}
If the equality $\tilde W'(q)=
W(q)/ 2 MT$ is fulfilled for the function $W({\bf q,p)}$, we arrive at the usual Einstein
relation for the coefficients $A_\alpha$ and $B_{\alpha \beta}$
\begin{equation}
M T A_\alpha({\bf p}) =  p_\alpha B \label{DC11b}
\end{equation}

Let us check this relation for Boltzmann collisions which are
described by the PT-function $W({\bf q, p)}=w_B({\bf q, p})$ [11]:
\begin{eqnarray}
w_B({\bf q, p})=\frac{2\pi}{\mu^2 q} \int_{q/2\mu}^\infty du\,u\,
\frac{d \sigma}{do} \left[\arccos \, (1-\frac{q^2}{2\mu^2 u^2}), u
\right] f_b (u^2+ v^2-{\bf q \cdot v} /\mu), \label{DC12b}
\end{eqnarray}
where (${\bf p}=M{\bf v}$), the quantities $d \sigma (\chi,u)/ do\, $, $\mu$
and $f_b$ are, respectively, the scattering differential cross-section, the mass and distribution function
for light particles. The arguments $\chi$ and $u$ in the cross-section $d
\sigma/do\,$ are the angle of the particle scattering in the center-of-
mass coordinate system and the velocity of light particles
before collision. In Eq.~(\ref{DC12b}), we took into account the
approximate equalities for light and heavy
particle scattering, $q^2 \equiv (\triangle {\bf p})^2=p'^2(1-cos\theta)$ and
$\theta\simeq \chi$ (where $p'=\mu u$ is the light
particle momentum before collision). For the equilibrium Maxwellian
distribution $f_b^0$, the equality $\tilde W'(q)= W(q)/ 2 MT$ is
evident and we arrive at the ordinary Fokker-Planck equation in
velocity space with the constant diffusion $D \equiv B /M^2$ and
friction $\beta \equiv B/MT=DM/T$ coefficients which satisfy the
Einstein relation.

The problem of the determination of the $W$-function and the respective Fokker-Planck
equation for the situations close to equilibrium are discussed in detail in the review [22]. However, even for quasi-equilibrium regimes, where long tails of the PTF-functions are
absent, the consideration in [22] is restricted to the hard sphere interaction. The non-equilibrium forms of
the $W$-function and the drastic changes in the Fokker-Planck equation structure for these situations
were not considered, as well as the specific situations, when the long tail of the $W$-function exists. Therefore, anomalous diffusion in [22] is absent. As is easy to see, in the case of the
hard-sphere cross-section and the equilibrium Maxwellian distribution, $f_b$ Eq.~(\ref{DC12b}) leads to
the same result for the $W_B$ function as has been discussed in [22].

For some non-equilibrium (stationary or non-stationary) states, the PTF can have a long
tail as a function of $q$. In this case, we have to derive a generalization of the
Fokker-Planck equation in spirit of the consideration [1,11] for
coordinate case. The necessity of this derivation arises since the diffusion and friction
coefficients in the form Eqs.~(\ref{DC6b}),(\ref{DC10b}) diverge
in the limit of large $q$, if the kernels of the functions $W$, $\tilde W'$ exhibit the asymptotic behavior,
$W(q)\sim 1/q^\alpha$ with $\alpha\leq s+2$ and (or) $\tilde
W'(q)\sim 1/q^\beta$ with $\beta \leq s+2$.

Let us substitute the expansions for $W$ (as an
example, we choose $s=3$, the arbitrary $s$ can be considered by
the similar way) into Eq.~(\ref{DC2b}). With an accuracy up to $({\bf q \cdot
p})^2$, we find
\begin{eqnarray}
\frac{df_g({\bf p},t)}{dt}=\int d{\bf q} \{f_g({\bf p+q}, t)[W(q)+\tilde W'(q) \,({\bf q \cdot p})+\nonumber\\
\frac{1}{2}\tilde W''(q) ({\bf q \cdot p})^2 +q^2 \tilde
W'(q)]-f_g({\bf
p},t)[W(q)+\tilde W'(q) \,({\bf q \cdot p})+\frac{1}{2}\tilde
W''(q) ({\bf q \cdot p})^2]\} \label{DC13b}
\end{eqnarray}

After the Fourier-transformation $f_g ({\bf s},t)=\int \frac{d{\bf
p}}{(2\pi)^3} exp(i{\bf p r})f_g ({\bf p},t)$, Eq.~(\ref{DC13b})
reads
\begin{eqnarray}
\frac{df_g({\bf s},t)}{dt} = A(s)f_g ({\bf s},t)+ B_\alpha
(s)\frac{\partial f_g ({\bf s},t)}{\partial {\bf s}_\alpha}
+C_{\alpha\beta}(s)\frac{\partial^2 f_g ({\bf s},t)}{\partial {\bf
s}_\alpha
\partial {\bf
s}_\beta}\label{DC16b}
\end{eqnarray}
It should be noted that the variable ${\bf s}_\alpha$,
arisen in  Eq.~(\ref{DC16b}) has the dimension ${\bf
p}_\alpha^{-1}$ and is a formal variable for the
Fourier-transformation.

The coefficients in Eq.~(\ref{DC16b}) are given by
\begin{eqnarray}
A({\bf s})= \int d{\bf q} [exp(-i{\bf(q s)})-1]W(q) = 4\pi \int_0^\infty
dq q^2 \left[\frac{sin\, (q s)}{qs}-1\right]W(q) \label{DC17b}
\end{eqnarray}
\begin{eqnarray}
B_\alpha({\bf s})\equiv s_\alpha B(s);\;B(s)=-\frac{i}{s^2} \int d{\bf q}
{\bf q s} [exp(-i{\bf(q s}))-1]  \tilde W'(q)= \nonumber\\
\frac{4\pi}{s^2} \int_0^\infty dq q^2 \left[cos\, (q s)-\frac{sin
(q s)}{q s}\right]\tilde W'(q) \label{DC19b}
\end{eqnarray}
\begin{eqnarray}
C_{\alpha\beta} (s)\equiv s_\alpha s_\beta C(s)= -\frac{1}{2}\int
d{\bf q} q_\alpha q_\beta [exp(-i{\bf(q s}))-1]  \tilde
W''(q)\label{DC20b}
\end{eqnarray}
\begin{eqnarray}
C(s)=-\frac{1}{2 s^4} \int d{\bf q} {\bf (q s)^2} [exp(-i({\bf q
s}))-1] \tilde W''(q)=\nonumber\\\frac{2\pi}{s^2} \int_0^\infty dq
q^4 \left[\frac{2 sin (q s)}{q^3 s^3}-\frac{2 cos\, (q s)}{q^2
s^2}-\frac{sin (qs)}{qs}+\frac{1}{3}\right]\tilde W''(q)
\label{DC21b}
\end{eqnarray}

For the isotropic function $f({\bf s})=f(s)$, one can rewrite
Eq.~(\ref{DC16b}) as
\begin{eqnarray}
\frac{df_g(s,t)}{dt} = A (s) f_g (s,t)+ B(s)s \frac{\partial f_g
(s,t)}{\partial s} +C(s) s^2 \frac{\partial^2 f_g(s,t)}{\partial
s^2} \label{DC22b}
\end{eqnarray}

In the case where the PT-function $W(q)$ and the functions $\tilde W'(q)$
and $\tilde W''(q)$ strongly decrease for large values of $q$, the
exponents under the integrals in the functions $A(s)$, $B(s)$, and
$C(s)$ can be expanded as
\begin{eqnarray}
A(s)\simeq-\frac{s^2}{6}\int d{\bf q}\, q^2 W(q);\; B(s)\simeq -
\frac{1}{3} \int d{\bf q} \, q^2 \tilde W'(q);\;C(s)\simeq 0.
\label{DC23b}
\end{eqnarray}

Then, the simplified kinetic equation in velocity space
PT-function (which is non-equilibrium in the general case) is written as
\begin{eqnarray}
\frac{df_g(s,t)}{dt} = A_0 s^2 f_g(s)+ B_0 s \frac{\partial f_g
(s)}{\partial s} ,\label{DC25b}
\end{eqnarray}
where $A_0\equiv -1/6 \int d{\bf q}\, q^2 W(q)$ and $B_0\equiv
-1/3 \int d{\bf q} \, q^2 \tilde W'(q)$ is defined by Eq.~(\ref{DC23b}).

For $C(s)=0$ the stationary solution of Eq.~(\ref{DC22b}) is given by
\begin{eqnarray}
f_g(s)=C exp\,\left[-\int_0^s ds'\frac{A(s')}{s' B(s')}\right] =C
exp\,\left[-\frac{A_0 s^2}{2 B_0}\right] \label{DC26b}
\end{eqnarray}

The respective normalized stationary momentum distribution is written as
\begin{eqnarray}
f_g(p)=\frac{N_g B_0^{3/2}}{(2 \pi A_0)^{3/2}}exp \,[-\frac{B_0
p^2}{2 A_0}]\label{DC28b}
\end{eqnarray}
Therefore, the constant in  Eq.~(\ref{DC26b}) is $C=N_g$, where $N_g$
is the density of heavy particles. Equation (\ref{DC25b}) and the obtained
distribution are the generalization (for non-equilibrium situations) of the standard Fokker-Planck equation to the case of normal diffusion in velocity space (see,e.g., [21]). The characteristic feature  of these physical situations is the existence of the fixed (prescribed) kernel
$ W({\bf q,p})$ which is defined, e.g., by a certain non-Maxwellian
distribution of small particles $f_b$. To show this in other
way, let us make the Fourier transformation of Eq.~(\ref{DC16b})
with $C=0$ and respective coefficients $A$ and $B_\alpha$,
\begin{eqnarray}
\frac{df_g({\bf p},t)}{dt} = - A_0 \frac{\partial^2 f_g({\bf p},t)}{\partial p^2}
- B_0 \frac{\partial (p_\alpha f_g
({\bf p},t))}{\partial p_\alpha} ,\label{DC29b}
\end{eqnarray}
Therefore, we arrive at the Fokker-Planck-type equation with the
friction coefficient $\beta\equiv-B_0$ and diffusion coefficient
$D=-A_0/M^2$. In general, these coefficients (Eq.~(\ref{DC23b})) do
not satisfy the Einstein relation.

In the case of equilibrium (e.g., $f_b=f_b^0$, see
above) for the $W$-function the equality $\tilde W'(q)=W(q)/2M T_b$ is fulfilled. Then,
with the necessary accuracy of the order $\mu/M$, we find $A(s)/s B(s)\equiv A_0/B_0 =M T_b$. In this
case, the Einstein relation between the diffusion and friction
coefficients $D=\beta T/M$ exists, and the standard Fokker-Planck
equation is valid.

In fact, the approximation $C(s)\simeq 0$ is practically always
applicable due to the small parameter (e.g., $\mu/M$ for the
Boltzmann-type PTF, see below, Sec.III). Therefore, the general kinetic
equation (\ref{DC22b}) for the Fourier-transform of the velocity
distribution function takes the form
\begin{eqnarray}
\frac{df_g({\bf s},t)}{dt} = A({\bf s})f_g ({\bf s})+ B_\alpha
({\bf s})\frac{\partial f_g ({\bf s},t)}{\partial {\bf s}_\alpha},
\label{DC29c}
\end{eqnarray}
where the coefficients $A({\bf s})$ and $B_\alpha({\bf s})$ are determined by (\ref{DC17b}),(\ref{DC19b}), respectively.

\section{The models of anomalous diffusion in $V$ - space}

Now we can calculate the coefficients for the models of anomalous
diffusion. We will also estimate for the function $W''$ and the tensor $C_{\alpha,\beta}$; although,
at the end we neglect these terms due to the small transfer momentum in the collision process.

At first, we calculate the simple model, i.e., the system of hard spheres
with different masses $m$ and $M\gg m$, $d\sigma/do=a^2/4$.
Let us suppose that light particles in the model under consideration are described by the prescribed stationary distribution
$f_b=n_b \phi_b /u_0^3$ (where $n_b$, $\phi_b$, and $u_0$ are,
respectively, the density, non-dimensional distribution, and
characteristic velocity of light particles). The integration variable is defined by the equality $\xi \equiv
(u^2+ v^2-{\bf q\cdot v} /\mu)/u_0^2$,
\begin{eqnarray}
W_a ({\bf q, p})= \frac{n_b a^2 \pi}{2\mu^2 u_0 q}
\int^\infty_{(q^2/4\mu^2+v^2-{\bf{q\cdot v}}/\mu)/u_0^2} d\xi
\,\cdot \phi _b (\xi).\label{DC30b}
\end{eqnarray}

If the distribution $\phi_b(\xi)=1/\xi^\gamma$ ($\gamma>1$)
has a long tail, we get
\begin{eqnarray}
W_a({\bf q, p})= \frac{n_b a^2 \pi}{2\mu^2 u_0
q}\frac{\xi^{1-\gamma}}{(1-\gamma)}|_{\xi_0}^\infty= \frac{n_b a^2
\pi}{2\mu^2 u_0 q}\frac{\xi_0^{1-\gamma}}{(\gamma-1)}
,\label{DC31b}
\end{eqnarray}
where $\xi_0\equiv (q^2/4\mu^2+v^2-{\bf{q\cdot v}}/\mu)/u_0^2$.

For the case $p=0$, $\xi_0 \rightarrow \tilde \xi_0
\equiv q^2/4\mu^2 u_0^2$, and we arrive at the expression for the
anomalous PT-function $W \equiv W_a$,
\begin{eqnarray}
W_a ({\bf q, p=0})=\frac{n_b a^2
\pi}{2^{3-2\gamma}(\gamma-1)\mu^{4-2\gamma}
u_0^{3-2\gamma}q^{2\gamma-1}}\equiv \frac{C_a}{q^{2\gamma-1}}
.\label{DC32b}
\end{eqnarray}

To determine the transport process the structure and the
kinetic equation in velocity space, one should also find the functions
$\tilde W'(q)$ and $\tilde W''(q)$.

To find the functions $\tilde W'({\bf q, p})$ and $\tilde W''({\bf q, p})$ for $p\neq 0$, we
use full-value $\xi_0\equiv (q^2/4\mu^2+p^2/M^2-{\bf{q\cdot
p}}/M \mu)/u_0^2$ and the derivatives of these functions on ${\bf q \cdot p}$ at
$p=0$, $\xi'_0=-1/M \mu u_0^2$, and $\xi''_0=0$. Then
\begin{eqnarray}
\tilde W'({\bf q, p})\equiv \frac{n_b a^2 \pi}{2M \mu^3 u^3_0
q}\xi_0^{-\gamma};
 \;\; \; \tilde W''({\bf q, p})\equiv \frac{n_b
a^2 \pi\gamma}{2M^2 \mu^4 u^5_0 q}\xi_0^{-\gamma-1} \label{DC34b}
\end{eqnarray}
Therefore, for $p=0$ ($\xi_0 \rightarrow \tilde \xi_0$) we obtain
the functions
\begin{eqnarray}
\tilde W'(q)\equiv \frac{(4 \mu^2u_0^2)^\gamma n_b a^2 \pi}{2M
\mu^3 u^3_0 q^{2\gamma+1}};
 \;\; \; \tilde W''(q)\equiv \frac{(4 \mu^2u_0^2)^{\gamma+1} n_b
a^2 \pi\gamma}{2M^2 \mu^4 u^5_0 q^{2\gamma+3}} \label{DC35b}
\end{eqnarray}

The function $A(s)$, according to Eq.~(\ref{DC17b}), is given by
\begin{eqnarray}
A(s)\equiv 4\pi \int_0^\infty d q q^2 \left[\frac{sin\, (q
s)}{q s}-1\right]W(q)= 4\pi C_a
 \int_0^\infty d q
\frac{1}{q^{2\gamma-3}} \left[\frac {sin(q s)}{q s}-1\right]
\label{DC33b}
\end{eqnarray}

Comparing the reduced equation (see below) in velocity space
with diffusion in coordinate space (see the Appendix), we
can establish the correspondence ($2\gamma-1\leftrightarrow\alpha$ and $W(q)=C/q^{2\gamma-1}$).
This means that the convergence of the integral on the right
side of Eq.~(\ref{DC33b}) (3d case) is provided if $3<2\gamma-1<5$
or $2<\gamma<3$. The inequality $\gamma<3$ provides the
convergence for small values of the variable $q$ ($q\rightarrow0$), and the inequality
$\gamma>2$ provides the convergence for $q\rightarrow\infty$.

We now establish the conditions of the convergence of the integrals
for $B(s)$ and $C(s)$ as

\begin{eqnarray}
B(s)= \frac{4\pi}{s^2} \int_0^\infty dq q^2 \left[cos\, (q
s)-\frac{sin (q s)}{q s}\right]\tilde W'(q) \label{DC36b}
\end{eqnarray}

The convergence of the function $B(s)$ exists for small values of $q$ if $\gamma<2$ and for
large values of $q$\, ($q\rightarrow\infty$ for $\gamma>1/2$.

Finally, the convergence for the function $C(s)$  is defined by the equalities
$\gamma<2$ for small values of $q$ and $\gamma>1$ for large values of $q$
\begin{eqnarray}
C(s)=\frac{2\pi}{s^2} \int_0^\infty dq q^4 \left[\frac{2 sin (q
s)}{q^3 s^3}-\frac{2 cos\, (q s)}{q^2 s^2}-\frac{sin
(qs)}{qs}+\frac{1}{3}\right]\tilde W''(q) \label{DC37b}
\end{eqnarray}
Therefore, in the case under consideration, the convergence of the functions $A$, $B$, and $C$ for large
large values of $q$ is defined only by the convergence of the function $A$, which means $\gamma>2$. To
provide convergence for small values of $q$, it is sufficient to provide
convergence for the functions $B(s)$ and $C$, which means $\gamma<2$. Therefore, for
the purely power behavior of the function $f_b(\xi)$, the
simultaneous convergence of the coefficients $A$, $B$ and $C$
cannot be provided. However, in real physical models, the convergence of the
coefficients in the integration region $q\rightarrow 0$ is
always provided, e.g., by a finite value of $v$ (see
Eq.~(\ref{DC31b})) or due to the non-power behavior of the PT-function $W$ for small values of
$q$  (compare with the examples of
anomalous diffusion in coordinate space [1]). Therefore, in the
model under consideration, the "anomalous diffusion in velocity
space"  for the power behavior of $W(q)$, $\tilde W'(q)$ and
$\tilde W''(q)$ at large values of the variable $q$ exists for the asymptotic behavior of the PT-function
$W(q\rightarrow\infty)\sim 1/q^{2\gamma-1}$ if the inequality
$\gamma>2$ is fulfilled. In this case the integrals for the coefficients
$A(s)$,$B(s)$ and $C(s)$ converge. At the same time, the
expansion of the exponential function in
Eqs.~(\ref{DC17b})-(\ref{DC21b}) under the integrals, which leads
to the Fokker-Planck type kinetic equation, is invalid for the
power-type kernels $W(\bf {q, p)}$.

Now let us consider the more general model for which we will not
connect the functions $W(q)$, $\tilde W'(q)$, and $\tilde W''(q)$
with a concrete form of $W({\bf q, p})$ which is in general unknown. In this case one can
suggest that the functions have the independent one from
another power-type $q$-dependence.

As an example, this dependence can be taken as the power-type one for
three functions $W(q)\equiv a / q^\alpha$, $\tilde W'(q)\equiv b
/q^\beta$, and $\tilde W''(q)\equiv c /q^\eta$ , where $\alpha$,
$\beta$ and $\eta$ are independent and positive. Then, as follows
from the above consideration, the convergence of the function $W$
exists if \,$5>\alpha>3$ (for asymptotically small and large $q$,
respectively). For the function $\tilde W'(q)$, the convergence
condition for asymptotically small and large values of $q$ is provided if $5>\beta>2$,
respectively.
Finally, for the function $\tilde W''(q)$, the
convergence condition is $7>\eta>5$ (for asymptotically small and
large values of $q$, respectively).

For this example, the kinetic equation Eq.~(\ref{DC16b}) reads
\begin{eqnarray}
\frac{df_g({\bf s},t)}{dt} = P_0 s^{\alpha-3} f({\bf s},t)+
s^{\beta-5} P_1 s_i\frac{\partial}{\partial s_i} f({\bf s},t)+
s^{\eta-7}P_2 s_i s_j \frac{\partial^2}{\partial s_i
\partial s_j}f({\bf s},t), \label{DC38b}
\end{eqnarray}
where
\begin{eqnarray}
P_0= 4\pi a \int_0^\infty d\zeta \zeta^{2-\alpha}
\left[\frac{sin\, \zeta}{\zeta}-1\right] \label{DC39b}
\end{eqnarray}
\begin{eqnarray}
P_1= 4\pi b \int_0^\infty d\zeta \zeta^{2-\beta} \left[cos\,
\zeta-\frac{sin \zeta}{\zeta}\right] \label{DC40b}
\end{eqnarray}
\begin{eqnarray}
P_2=4\pi c \int_0^\infty d\zeta \zeta^{4-\eta} \left[\frac{ sin
\zeta}{\zeta^3}-\frac{ cos\,\zeta}{\zeta^2}-\frac{sin
\zeta}{2\zeta}+\frac{1}{6}\right] \label{DC41b}
\end{eqnarray}

As is easy to show, $P_0=-|P_0|\,sgn \,a<0$ and $P_1=-|P_1|\,sgn\, b$.
Taking into account the isotropy in $s$-space we can rewrite
Eq.~(\ref{DC38b}) in the form
\begin{eqnarray}
\frac{df_g(s,t)}{dt} = P_0 s^{\alpha-3} f(s,t)+ s^{\beta-4} P_1
\frac{\partial}{\partial s} f(s,t)+ s^{\eta-5}P_2
\frac{\partial^2}{\partial s^2}f(s,t), \label{DC42b}
\end{eqnarray}

Naturally, Eqs.~(\ref{DC38b}),(\ref{DC42b}) can be formally
rewritten in momentum (or in velocity) space in terms of fractional
derivatives of various orders. Therefore, as is easy to see, for
the purely power behavior of the functions  $W(q)$, $\tilde W'(q)$
and $\tilde W''(q)$ the solution with the convergent coefficients
exists for the values of the powers in the intervals mentioned above. In the
case under consideration the universal type of the anomalous diffusion in velocity space exists if $5>\alpha>3$,
$5>\beta>2$ and $7>\eta>5$. This takes place even in the cases, when
the functions $W(q)$, $\tilde W'(q)$ and $\tilde W''(q)$ have not
cut-off for small values of $q$. Surely, the general
description is also valid for more complicated functions $W$,
$W'$, and $W''$, characterized by the non-power $q$-dependence for small values of $q$.

Now let us take into account an important circumstance; in general, there is
a small parameter $~\mu/M$ in the problem under consideration , which can simplify the description of velocity diffusion.
As is easy to see, e.g., based
on the particular cases (see Eq.~(\ref{DC35b})) for the convergent kernels of anomalous transport, the term
with the second derivative $\tilde W''$ in general equations (\ref{DC16b}),(\ref{DC22b}) for the distribution $f_g({\bf p})$ is small (in comparison with the term with the second derivative. The similar smallness is observed for the case of normal diffusion in velocity space.
This smallness is of the order of the small ratio $\mu/M$ of particle masses.

Therefore, for the most physically important kernels describing anomalous velocity diffusion, the term with the
second space derivative can be omitted and the non-stationary \emph{general diffusion equation} is given by
\begin{eqnarray}
\frac{df_g({\bf s},t)}{dt} = A(s)f_g ({\bf s},t)+ B_\alpha
(s)\frac{\partial f_g ({\bf s},t)}{\partial{\bf s}_\alpha}
\label{DC47a}
\end{eqnarray}
or, for the isotropic case,
\begin{eqnarray}
\frac{df_g(s,t)}{dt} = A (s) f_g(s,t)+ B(s)s \frac{\partial
f_g(s,t)}{\partial s}  \label{DC47b}
\end{eqnarray}
In the case of the purely power behavior, $W(q)=a/q^\alpha$ and
$W'(q)=b/q^\beta$, we have, as above, $A(s)=P_0 s^{\alpha-3}$ and
$sB(s)=P_1 s^{\beta-4}$ (with the inequalities $5>\alpha>3$ and
$5>\beta>2$). The stationary solution of Eq. (\ref{DC47b}) (see,
also (\ref{DC26b})) for the case under consideration is written as
\begin{eqnarray}
f^{St}_g (s)= C exp\,\left[-\int^s ds'\frac{A(s')}{s' B(s')}\right]=C
exp\,\left[-\frac{P_0 s^{\alpha-\beta+2}}{P_1 (\alpha-\beta+2)}\right],
\label{DC47b1}
\end{eqnarray}
where $5>\alpha-\beta+2>0$.

To find the solution in the isotropic non-stationary case, Eq.~(\ref{DC47b}) should be written as
\begin{eqnarray}
\frac{d X (s,t)}{dt}-B(s)s \frac{\partial}{\partial s} X (s,t) = A
(s), \label{DC48b}
\end{eqnarray}
where $X(s,t)\equiv ln f_g(s,t)$. The general non-stationary
solution of this equation can be written as the sum of the general
solution of the homogeneous equation $Y(s,t)$ (Eq.~(\ref{DC48b}),
where the function $A(s)$ is taken zero),
\begin{eqnarray}
Y (s,t) = \Phi (\xi), \,\, \xi \equiv t+\int^s_{s_0}
ds'\frac{1}{s' B(s')}.  \label{DC49b}
\end{eqnarray}
Here $\Phi$ is the arbitrary function of the variable $\xi$. The particular solution
$Z(s,t)$ of the inhomogeneous equation Eq.~(\ref{DC48b}) reads
\begin{eqnarray}
Z (s,t)\equiv f^{St}_g (s)= -\int^s_{s_0} ds'\frac{A(s')}{s'
B(s')} \label{DC50b}
\end{eqnarray}
Therefore, we find
\begin{eqnarray}
f_g(s,t) = exp\, [X(s,t)] \equiv exp \left[ Y+Z
\right]=L(\xi)f^{St}_g (s), \label{DC51b}
\end{eqnarray}
where $L(\xi)$ is the arbitrary function of $\xi$, which should be
found from the initial condition $f_g(s,t=0)\equiv \phi_0(s)$.

The variable $\xi\equiv \xi (s,t)$ equals
\begin{eqnarray}
\xi (s,t) =t+ \int^s_{s_0} ds'\frac{1}{s'
B(s')}=t+\frac{s^{5-\beta}}{P_1 (5-\beta)}+c,  \label{DC53b}
\end{eqnarray}
where $c$ is the arbitrary constant which can be omitted due to the presence of the arbitrary function
$L$.

The general non-stationary solution for the case under consideration reads
\begin{eqnarray}
f_g(s,t)=L \left(t+\frac{s^{5-\beta}}{P_1 (5-\beta)}\right) \, exp
\left[-\frac{P_0 s^{\alpha-\beta+2}}{P_1 (\alpha-\beta+2)}\right].
\label{DC54b}
\end{eqnarray}
The inequalities for the combinations of the above coefficients are $5>\alpha-\beta+2>0$, $3>5-\beta>0$.
The unknown function $L$ can be found from Eq.~(\ref{DC54b}) and
the initial condition $f_g(s,0)\equiv \phi_g(s)$:
\begin{eqnarray}
L \left(\frac{s^{5-\beta}}{P_1 (5-\beta)}\right) \, exp
\left[-\frac{P_0 s^{\alpha-\beta+2}}{P_1
(\alpha-\beta+2)}\right]=\phi_g(s), \label{DC55b}
\end{eqnarray}
The function $\phi_g(s)=\int d^3\, p \,exp (i{\bf p s})\, f_g(p, t=0)$
is the Fourier-component of the initial distribution in momentum
space. Using the notation $\zeta\equiv s^{5-\beta}/[P_1
(5-\beta)]$ (which means
$s(\zeta)\equiv[P_1(5-\beta)\zeta]^{1/(5-\beta)}$), we find
\begin{eqnarray}
L(\zeta) =\phi_g[s(\zeta)]\, exp \left\{\frac{P_0
[s(\zeta)]^{\alpha-\beta+2}}{P_1 (\alpha-\beta+2)}\right\},
\label{DC55b1}
\end{eqnarray}
Therefore, the time-dependent solution is
\begin{eqnarray}
f_g(s,t)=\phi_g[s(\zeta+t)]\, exp \left\{\frac{P_0
[s(\zeta+t)]^{\alpha-\beta+2}}{P_1 (\alpha-\beta+2)}\right\}\, exp
\left[-\frac{P_0 s^{\alpha-\beta+2}}{P_1 (\alpha-\beta+2)}\right],
\label{DC56b}
\end{eqnarray}
where we have to express the value
$s(\zeta+t)\equiv[P_1(5-\beta)(\zeta+t)]^{1/(5-\beta)}$ as the evident
function of variables $s, t$:
\begin{eqnarray}
s(\zeta+t)\equiv[P_1(5-\beta)(\zeta+t)]^{1/(5-\beta)}\equiv[s^{5-\beta}+P_1(5-\beta)t]^{1/(5-\beta)},
\label{DC57b}
\end{eqnarray}
or, finally,

\begin{eqnarray}
f_g(s,t)=\phi_g\left([s^{5-\beta}+P_1(5-\beta)\,t]^{1/(5-\beta)}\right)\,
exp \left\{\frac{P_0\,
[s^{5-\beta}+P_1\,(5-\beta)\,t]^{(\alpha-\beta+2)/(5-\beta)}-P_0
\,s^{\alpha-\beta+2}}{P_1\, (\alpha-\beta+2)}\right\}.
\label{DC56b2}
\end{eqnarray}
It should be emphasized that the real
solution for the fractional powers
$1/(5-\beta)$ and (or) $(\alpha-\beta+2)/(5-\beta)$  exists only if $P_1>0$. The limit $t \rightarrow \infty$
for the solution  can
be identical to the stationary solution only for specific initial conditions. For these cases, the stationary solution can be non-equilibrium.

For the power dependence of the functions $W(q)$ and $\tilde
W'(q)$, the equation (\ref{DC47a}) can be formally written in terms of
fractional derivatives,
\begin{eqnarray}
\frac{df_g({\bf p},t)}{dt} = P_0 D^\nu f_g({\bf p},t)-P_1 (3+\gamma)D^\gamma f_g({\bf p},t)+P_1 p_\alpha D_\alpha^{\gamma+1}
f_g ({\bf p},t), \label{DC56b3}
\end{eqnarray}
where $\nu\equiv \alpha-3$, $\gamma\equiv \beta-5$ and
$D_\alpha^{\gamma+1}f_g ({\bf p},t) \equiv i \int\, d^3 \,s \,exp
(-i{\bf ps}) )s_\alpha s^\gamma f_g ({\bf s},t)$.

The particular case of the collision term  for anomalous diffusion in velocity space, considered in [17],
is derived in Appendix II on the basis of the general equations (\ref{DC47a}),(\ref{DC56b}).

\section{Diffusion model based on Boltzmann collisions with drift}

Let us consider the simplest case of the non-equilibrium but
stationary distribution $f_b$, i.e., the shifted velocity
distribution.

The evident generalization of the PT-function $w_B^d({\bf q, p})$ for this case (characterizing by the drift velocity ${\bf u_d}$ in the distribution function $f_b$) is given by
\begin{eqnarray}
w_B^d({\bf q, p, u_d})= \frac{2\pi}{\mu^2 q} \int_{q /2\mu}^\infty d u
u\,\cdot \frac{d \sigma}{do} \left[\arccos \, (1-\frac{q^2}{2
\mu^2 u^2}), u \right] \times \nonumber\\
f_b (u^2+ ({\bf v-u_d})^2-{\bf q} \cdot {\bf (v-u_d)} /\mu)
,\label{DB1}
\end{eqnarray}
where ${\bf p}=M {\bf v}$.
Again, as in Section II, to find the coefficients in the
kinetic equation, let us use the difference
between the velocities of light and heavy particles. At the same time, the drift velocity $u_d$,
generally speaking, is not small in comparison with the current characteristic velocities
$u$ and $q/\mu$ of small particles.

To calculate the function $A_\alpha$, we have take into
account that the scalar function $w_B^d({\bf q, p,
u_d})$ can in general be written in the form $W({\bf q,p,u_d})=W(q,p,u_d,
l,\xi,\eta)$ (here $l \equiv ({\bf q \cdot p_d}$)) and expanded on
$\xi \equiv ({\bf q\cdot p})$ and $\eta\equiv (M {\bf u_d \cdot
p})\equiv ({\bf p_d \cdot p})$. In fact, it is the expansion in the
velocity ${\bf v}={\bf p}/M$, which is small in comparison with other
characteristic velocities $q/\mu$, $u$, and, in the general case, $u_d$. As shown
above (for the case $u_d=0$), to arrive at the simple and solvable
equation for the distribution function $f_g$, taking into account the smallness
of $v$ in comparison with characteristic velocities and $v^2$
in comparison with $({\bf q \cdot v}/\mu$)), we have approximate
the function $W(q,p,u_d, l,\xi,\eta) \simeq W(q,u_d, l,\xi,\eta)$,
because we are interested mainly in high values of $q$ for
anomalous transport. At the same time, after this approximation and expansion on $\xi$ and $\eta$ (for $u_d \neq 0$),  we can arrive (for the special case of kernels with the purely power-type $q$-dependence for small $q$) at the divergence of some coefficients of the diffusion equation. A similar situation takes place also
for anomalous diffusion in coordinate space. This divergence is absent for the realistic PT-functions,
which provide the cut-off for small values of $q$. This cut-off has the physical nature and is not related to the approximation $W({\bf q,p,u_d})\equiv
W(q,p,u_d, l,\xi,\eta) \simeq W(q,v=0,u_d, l,\xi,\eta) W(q,u_d, l,\xi,\eta)$.

Let us expand the PT-function $W(q,p, u_d, l,\xi,\eta)$ in $\xi$ and $\eta$ similarly to the more
simple case (\ref{DC7b}),
\begin{eqnarray}
W({\bf q,p,u_d})= W(q,p, u_d, l,\xi,\eta)\simeq
W_0(q,p, u_d,l)+\partial W/\partial \xi\mid_{\xi, \eta=0} \xi +
\partial W/\partial \eta\mid_{\xi, \eta=0}  \eta
+\nonumber\\
\frac {1}{2}\partial^2 W/\partial \xi^2 \mid_{\xi, \eta=0}
\xi^2+\frac {1}{2}\partial^2 W/\partial \eta^2 \mid_{\xi, \eta=0}
\eta^2+
\partial^2 W/\partial \xi \partial \eta \mid_{\xi, \eta=0}\xi \eta
,\label{DB2}
\end{eqnarray}
Here $W_0(q,p,u_d,l)\equiv W(q,p, u_d, l,\xi,\eta)\mid_{\xi, \eta=0}$.
Then, introducing the functions $V_1(q,p, u_d,l)=\partial W/\partial \xi\mid_{\xi,
\eta=0}, U_1(q,p,u_d,l)=\partial W/\partial \eta\mid_{\xi, \eta=0},
V_2(q,p,u_d,l)=\frac {1}{2}\partial^2 W/\partial \xi^2 \mid_{\xi,
\eta=0}, U_2(q,p,u_d,l)=\frac {1}{2}\partial^2 W/\partial \eta^2
\mid_{\xi, \eta=0}, W_2(q,p,u_d,l)=\partial^2 W/\partial \xi
\partial \eta \mid_{\xi, \eta=0}$ we can rewrite Eq.~(\ref{DB2})
in the form
\begin{eqnarray}
W(q,p, u_d, l,\xi,\eta) \simeq W_0(q,p,u_d,l)+V_1 ({\bf q\cdot p}) +\nonumber\\
U_1 ({\bf p_d \cdot p}) + V_2 ({\bf q\cdot p})^2+U_2 ({\bf p_d
\cdot p})^2+ W_2 ({\bf q\cdot p}) ({\bf p_d \cdot p})\simeq \nonumber\\ W_0(q,p=0,u_d,l)+V_1(q,p=0,u_d,l) ({\bf q\cdot p})+
U_1(q,p=0,u_d,l)({\bf p_d \cdot p}) , \label{DB3}
\end{eqnarray}
Finally, we set $p=0$ in the coefficients of Eq.~(\ref{DB3}) and omit the terms with the second order derivatives due to the existence of the small parameter (e.g., $~\mu/M$). This means only the terms of the order of $\sqrt{\mu/M}$ are essential in the expansion of $W({\bf q,p,u_d})$.
Let us calculate $W ({\bf q, p+q,u_d})$ taking into account the difference
between the values of the characteristic momenta ${\bf q, p+q,u_d}$ or the characteristic values
of velocities.

The appropriate expansion for the function $W ({\bf q, p+q,u_d})$ with the necessary accuracy is written as
\begin{eqnarray}
W ({\bf q, p+q, u_d})\simeq W ({\bf q, p, u_d})+\{q_{\alpha} \partial /\partial
p_{\alpha}+\frac{1}{2}q_\alpha q_{\beta} \frac{\partial^2
}{\partial p_ {\alpha} \partial p_{\beta}} \}W({\bf q,
p,u_d})\simeq \nonumber\\W_0(q,p=0,u_d,l)+V_1(q,p=0,u_d,l) ({\bf q\cdot p})+
U_1(q,p=0,u_d,l) ({\bf p_d \cdot p})+\nonumber\\
V_1(q,p=0,u_d,l) {\bf q}^2+
U_1(q,p=0,u_d,l) ({\bf q \cdot  p_d})
\label{DB3a}
\end{eqnarray}
or
\begin{eqnarray}
W ({\bf q, p+q, u_d})\simeq W_0(q,p=0,u_d,l)+\nonumber\\ V_1(q,p=0,u_d,l) [{\bf q\cdot (p+q)}]+
U_1(q,p=0,u_d,l) [{\bf p_d \cdot (p+q)}].
\label{DB4}
\end{eqnarray}
Then the collision term of the kinetic equation can be given by the formula
\begin{eqnarray}
\frac{df_g({\bf p},t)}{dt} = \int d{\bf q} \{[W_0(q,p=0,u_d,l)+V_1(q,p=0,u_d,l) ({\bf q\cdot p})+
U_1(q,p=0,u_d,l) ({\bf p_d \cdot p})\nonumber\\+
V_1(q,p=0,u_d,l) {\bf q}^2+U_1(q,p=0,u_d,l)({\bf q \cdot  p_d})]\nonumber\\ f_g({\bf p+q},t)- [W_0(q,p=0,u_d,l)+V_1(q,p=0,u_d,l)
({\bf q\cdot p})+\nonumber\\
U_1(q,p=0,u_d,l) ({\bf p_d \cdot p})] f_g({\bf p}, t)\}.
\label{DB5}
\end{eqnarray}
After the Fourier-transformation (\ref{DC13b}), the distribution function $f_g ({\bf s})=\int \frac{d{\bf
p}}{(2\pi)^3} exp(i{\bf p s})f_g ({\bf p},t)$
can be written in the form
\begin{eqnarray}
\frac{df_g({\bf s},t)}{dt}= \int d{\bf q}\{exp(-i{\bf q
s)}[W_0(q,p=0,u_d,l)-i V_1(q,p=0,u_d,l) ({\bf q\cdot\frac{\partial}{\partial{\bf s}}})-\nonumber\\
i U_1(q,p=0,u_d,l) ({\bf p_d \cdot \frac{\partial}{\partial{\bf
s}} })]f_g({\bf s},t)\nonumber\\ - [W_0(q,p=0,u_d,l)-i
V_1(q,p=0,u_d,l) ({\bf q\cdot \frac{\partial}{\partial{\bf s}}})-
i U_1(q,p=0,u_d,l) ({\bf p_d \cdot \frac{\partial}{\partial{\bf s}}})]f_g({\bf s},t)\}\nonumber\\
 \label{DB6}
\end{eqnarray}
Therefore, we arrive at the equation similar to Eq. (\ref{DC47a}), but with the evidently non-isotropic structure,
\begin{eqnarray}
\frac{df_g({\bf s},t)}{dt}=A_d({\bf s, p_d})f_g({\bf s},t)+{\bf
B}_d({\bf s, p_d})\frac{\partial}{\partial{\bf s}}f_g({\bf s},t),
\label{DB7}
\end{eqnarray}
where the coefficients are expressed as the integrals
\begin{eqnarray}
A_d({\bf s, p_d})=\int d{\bf q}[exp(-i{\bf q
s)}-1]W_0(q,p=0,u_d,l), \label{DB7a}
\end{eqnarray}
\begin{eqnarray}
{\bf B}_d({\bf s,p_d})=-i \int d{\bf q}[exp(-i{\bf q
s)}-1]\{V_1(q,p=0,u_d,l) {\bf q}+U_1(q,p=0,u_d,l){\bf
p}_d\}\equiv\nonumber\\{\bf s}B'_d({\bf s,p_d})+{\bf p_d}
B''_d({\bf s,p_d}). \label{DB8}
\end{eqnarray}
This equation will be analyzed in detail in the separate paper.

\section{Conclusions}

In this paper, the problem of anomalous diffusion in momentum
(velocity) space has been consistently considered. The new kinetic
equation for anomalous diffusion in velocity space has been derived for the
general case of the non-equilibrium probability transition
function.

It should be emphasized that these types of anomalous diffusion are not related to the
time dispersion (the non-local in time PT-function). Anomalous diffusion in velocity space, created by the non-local in time master equation has been considered in a few papers (see, e.g., [15],[16], [23],[24]) on the basis of the well-known Fokker-Planck kernel in $p$-space.

The similar problem with the non-local in time PT-functions, which at the same time have long $q$-tails, has been considered for diffusion in velocity space separately, based on the results obtained in this paper.

The evolution of the distribution function can lead to the stationary non-equilibrium
limit or be asymptotically time-dependent. The model of
anomalous diffusion in velocity space is described based on the
respective expansion of the kernel in the master equation. The
conditions of the convergence of the coefficients of the new kinetic
equation have been found for particular cases. The wide variety of
anomalous processes in velocity space exists, since two ($C$ is usually small) different coefficients enter the general diffusion equation even in the isotropic case.

The examples of anomalous diffusion for heavy particles,
based on the Boltzmann kernel with the prescribed equilibrium and
non-equilibrium distribution function for the light particles have been
studied. The PT-function for the Boltzmann kernel depends on the appropriate cross-section for scattering of
two types of colliding particles and on the prescribed velocity distribution for one sort (light in the case under consideration) particles. The particular case of the hard sphere interaction has been considered. The
Einstein relation is applicable only to the equilibrium PT-function
(hence, the equilibrium prescribed distribution of light
particles). In the general case, even for the stationary solution, the
Einstein relation between the diffusion and friction coefficients
is absent. For normal diffusion, the friction and diffusion
coefficients have been explicitly found for the non-equilibrium case. The
stationary and non-stationary (with an initial condition)
solutions  of the general kinetic equation have been found. In the
equilibrium case, the known Fokker-Planck equation is reproduced as
a particular case. We also have shortly reviewed anomalous
diffusion in coordinate space.

The kinetic equation obtained in this paper is applicable to various systems in physics (including plasma physics [25]), chemistry, and biology.

\section{Appendix}

\textbf{Appendix I. Diffusion in the coordinate space on the basis of
the master-type equation}

Let us consider diffusion in coordinate space on the basis of the
master equation which describes the balance of grains coming in
and out the point $r$ at the moment $t$. The structure of this
equation (see, e.g., [1,11]) is formally similar to the master equation in
momentum space, which has been derived above,
\begin{equation}
\frac{df_g({\bf r},t)}{dt} = \int d{\bf r'} \left\{W ({\bf r, r'})
f_g({\bf r', t}) - W ({\bf r', r}) f_g({\bf r},t) \right\}.
\label{DC1}
\end{equation}
Surely, for coordinate space, there is no conservation law
similar to that in momentum space. The probability transition
$W({\bf r, r'})$ describes the probability that a grain transfers
from the point ${\bf r'}$ to the point ${\bf r}$ per unit time. We
can rewrite this equation in the coordinates ${\bf u=r'- r}$ and
${\bf r}$ as
\begin{eqnarray}
\frac{df_g({\bf r},t)}{dt} = \int d{\bf \rho} \left\{W ({\bf u,
r+u}) f_g({\bf r+u},t) - W ({\bf u, r}) f_g({\bf r},t) \right\}.
\label{DC2}
\end{eqnarray}
Assuming that characteristic displacements are small, one can
expand Eq.~(\ref{DC2}) and arrive at the Fokker-Planck form of the
equation for the density distribution $f_g({\bf r},t)$,
\begin{equation}
\frac{df_g({\bf r},t)}{dt} = \frac {\partial}{\partial r_{\alpha}}
\left[A_\alpha ({\bf r}) f_g({\bf r},t) + \frac {\partial}
{\partial r_{\beta}} \left( B_{\alpha\beta}({\bf r}) f_g({\bf
r},t) \right)\right]. \label{DC3}
\end{equation}
The coefficients $A_\alpha$ and $B_{\alpha \beta}$ describing the
acting force and diffusion, respectively, can be written as the
functionals of the PT-function in coordinate space $W$. For the
dimension $s$, these coefficients have the form
\begin{equation}
A_\alpha({\bf r}) = \int d^s {\bf u} {\bf u}_\alpha W({\bf u, r})
\label{DC4}
\end{equation}
and
\begin{equation}
B_{\alpha\beta}({\bf r})= \frac{1}{2}\int d^s {\bf u} {\bf
u}_\alpha {\bf u}_\beta W({\bf u, r}). \label{DC5}
\end{equation}
For the isotropic case, the probability function depends on ${\bf
r}$ and the magnitude $u$ of the vector ${\bf u}$. For a homogeneous
medium, when the $r$-dependence of the PT-function is absent, the coefficient
$A_\alpha=0$, while the diffusion coefficient is the constant
$B_{\alpha\beta}=\delta_{\alpha\beta}B$. The constant $B$ is the integral
\begin{equation}
B = \frac{1}{2s}\int d^s {\bf u} u^2 W(u). \label{DC6}
\end{equation}

This consideration cannot be applied to the specific situations in
which the integral in Eq.~(\ref{DC6}) is infinite. In this case we
should examine the general transport equation (\ref{DC1}). Let us now
consider the problem for the homogeneous and isotropic
case, when the PT-function depends only on $u$. As shown in [11], after the
Fourier-transformation we arrive at the following representation of
Eq.~(\ref{DC1})
\begin{equation}
\frac{df_g({\bf k},t)}{dt} = \int d^s {\bf u}  \left[\exp (i {\bf
k u}) -1 \right]W(u) f_g({\bf k},t)\equiv X({\bf k})f_g({\bf
k},t), \label{DC7}
\end{equation}
where $X({\bf k})\equiv X(k)$. Let us assume the simple form of the
PT-function with the power dependence on the distance $W(\rho)=
C/u^\alpha$, where $C$ is a constant and $\alpha>0$. Such a
singular dependence is typical for the jump diffusion probability in
heteropolymers in solutions (see, e.g., [26], where different
applications of anomalous diffusion are considered based on
the fractional differentiation method). For the one-dimensional
case (s=1), we find
\begin{equation}
X(k)\equiv -4 \int_0^\infty d u \, sin^2
\left(\frac{k\,u}{2}\right)W(u)= - 2^{3-\alpha}C |k|^{\alpha-1}
\int_0^\infty d\zeta \frac{sin^2\zeta}{\zeta^\alpha}. \label{DC8}
\end{equation}
For $1<\alpha<3$, this function is finite and written as
\begin{equation}
X(k) = - \frac{C\, \Gamma[(3-\alpha)/2]\,|k|^{\alpha-1}}{2^\alpha
\sqrt\pi \, \Gamma(\alpha/2)(\alpha-1)} , \label{DC9}
\end{equation}
where $\Gamma$ is the Gamma-function. At the same time, the
integral in Eq.~(\ref{DC6}) for  such PT-functions is
infinite, since normal diffusion is absent.

The procedure considered for the simplest cases of the power
dependence of the PT-function is equivalent to the equation with
fractional space differentiation [10,26],
\begin{equation}
\frac{df_g(x,t)}{dt} = C\Delta^{\mu/2}f_g(x,t),\label{DC10}
\end{equation}
Here $\Delta^{\mu/2}$ is the fractional Laplacian, the linear
operator whose action on the function $f(x)$ in Fourier space is
described by the expression $\Delta^{\mu/2}f(x)=-(k^2)^{\mu/2}f(k)=-|k|^\mu
f(k)$. In the case considered above, $\mu\equiv(\alpha-1)$, where
$0<\mu<2$. For more general PT-functions, which (for arbitrary
values of the variable $u$) are not proportional to the power of $u$, the
method described above is also applicable, although the fractional
derivative method is not exist.

In the case of the purely power dependence of the PT-function, the non-stationary
solution for the density distribution describes the so-called
super-diffusion (or Levy flights). The solution of
Eq.~(\ref{DC10}) in the Fourier space reads
\begin{equation}
f_g(k,t) = \exp (-C|k|^\mu t),\label{DC11}
\end{equation}
which, in coordinate space, corresponds to the so-called symmetric
Levy stable distribution
\begin{equation}
f_g(x,t) = \frac {1}{(kt)^{1/\mu}} L\left[\frac{x}{(kt)^{1/\mu}};
\mu, 0 \right] \label{DC12}
\end{equation}
(here we use the canonical notation [27]).

In the general case it follows from Eq.~(\ref{DC7}) that
\begin{equation}
f_g(k,t) = C_1 \exp [X(k)t],\label{DC12a}
\end{equation}
with a certain constant $C_1$.

The consideration based on the PTF given in this Section
allows us to avoid the fractional differentiation method and to
consider more general physical situations of the non-power type of the
probability transitions. Let us consider a simple
example. Taking (for the one-dimensional case) the PT function
$W(u)$ in the form
\begin{equation}
W(u) = C \frac{1-\exp [-\sigma u^p]}{u^\alpha}, \label{DC14}
\end{equation}
with $p>0$, we arrive at the function $X(k)$:
\begin{equation}
X(k)= - 2^{3-\alpha}C |k|^{\alpha-1} \int_0^\infty d\zeta
\frac{\left\{1-\exp [-\sigma (2\zeta/|k|)^p\;]\right\}
sin^2\zeta}{\zeta^\alpha}\equiv -2^{3-\alpha}C
|k|^{\alpha-1}\,T(\sigma/|k|^p,\alpha). \label{DC15}
\end{equation}
It is easy to see that the function $T(\sigma/|k|^p,\alpha)$ is
finite for $1<\alpha<p+3$, since for small
distances at $p>0$ divergence is also suppressed for the powers
$\alpha>3$. A simple calculation for $\alpha=2$ and $p=1$ leads to
the following result which cannot be found by the usual
fractional differentiation method,
\begin{equation}
T(\sigma/|k|,2)= \frac{\pi}{2}-\arctan
(|k|/\sigma)+\frac{\sigma}{2|k|} \ln \left[1+k^2/\sigma^2\right].
\label{DC16}
\end{equation}
The asymptotic behavior of the function $X(k)$ for $k \rightarrow
0$ (or $\sigma\rightarrow\infty$) is similar, as follows from
Eq.~(\ref{DC16}), to the case $W(u)=C/u^\alpha$. For the case
under consideration, $\alpha=2$, the limit $X(k \rightarrow 0)
\rightarrow -\pi C k$. For large values of $k$
($k\rightarrow\infty$ or $\sigma\rightarrow 0$), we find
$X(k)\rightarrow (\sigma/k) \ln(k/\sigma)$.

In the general case, the universal behavior of the function $X(k)$ is
provided by the asymptotical properties of the PT-function for large
distances. The necessary condition for divergence of the coefficient is $1<\alpha<p+3$.

\textbf{Appendix II. Anomalous velocity diffusion for the specific
case B(s)=const, C(s)=0}

Let us now consider formally the particular case of anomalous
diffusion in velocity space, when the specific structure of PTF $W({\bf q, p})$
provides a rapid (let say, exponential) decrease in the functions
$\tilde W'(q)$ and $\tilde W''(q)$. Therefore, the exponential
function under the integrals in the coefficients $B(s)$ and $C(s)$
can be expanded, which means $B(s)=B_0$ and $C(s)\simeq 0$
respectively. At the same time, the function $W(q)\equiv
a/q^\alpha$ has a purely power dependence on $q$.

In this particular case, the kinetic equation
Eq.~(\ref{DC42b}) reads
\begin{eqnarray}
\frac{df_g({\bf s},t)}{dt} = P_0 s^{\alpha-3} f_g({\bf s},t)+ B_0
s_i\frac{\partial}{\partial s_i} f_g({\bf s},t), \label{A2a}
\end{eqnarray}
or, formally, in the momentum space
\begin{eqnarray}
\frac{df_g({\bf p},t)}{dt} = P_0 D^\nu f_g({\bf p},t)- B_0
\frac{\partial}{\partial p_i} [p_i f_g({\bf p},t)], \label{A2b}
\end{eqnarray}
where $\nu \equiv(\alpha-3)$ ($2>\nu>0$), and we introduced the
fractional differentiation operator $D^\nu f({\bf p},t)\equiv \int
d{\bf s} s^\nu exp(-i{\bf ps}) f({\bf s},t) $ in momentum
space. This is a very particular case of the general equations Eqs.~(\ref{DC29c}),(\ref{DC42b}). In this case the result is similar to the
particular phenomenological model [17]. The stationary solution of Eq.~(\ref{A2a})
is given by
\begin{eqnarray}
f_g(s) = C exp\;[-\frac {P_0 s^{\nu}}{\nu B_0}] \label{A2c}
\end{eqnarray}
\begin{eqnarray}
f_g(p) = C \int d^3 s exp(-i{\bf ps})exp\;[-\frac {P_0
s^{\nu}}{\nu B_0}]\equiv \frac{4\pi C}{p} \int^\infty_0  ds s sin
(ps) exp\;[-\frac {P_0 s^{\nu}}{\nu B_0}]\label{A2d}
\end{eqnarray}

As an example, let us choose the case $\nu=1$. Then, we find $f_g({\bf p})$
\begin{eqnarray}
f_g(p) = \frac{8\pi C'P_0}{B_0 [(p^2+4P_0^2/B_0^2)]^2}.\label{DF5a}
\end{eqnarray}
In the case $\nu=1$ the long tail of the distribution is proportional to ~$p^{-4}$ and the distribution $f_g(p)$ corresponds with the Cauchy-Lorentz distribution. Normalization of the distribution $f_g(p)$ leads to the value $C=n_g/(2\pi)^3$, where $n_g=N_g/V$ is the average density of particles undergoing diffusion in velocity space. A similar approach can be taken for other types of anomalous
diffusion in velocity space.

\section*{Acknowledgment}

This paper has been reported as the invited talk at the Kiev International Conference devoted to the 100th anniversary of N.N. Bogolyubov, in September 2009.

I appreciate to the Organizing Committee and the Chairman A.G. Zagorodny for invitation, as well as, many participants for the useful discussions. I am grateful to E.A. Allahyarov, M. Bonitz, W. Ebeling, A.M. Ignatov, M.Yu. Romanovsky, and I.M. Sokolov for many stimulating discussions of anomalous diffusion problems. I am grateful to M.S. Karavaeva for her permanent supporting interest to this work.

This study was  supported by the Netherlands Organization for Scientific Research (NWO) and the Russian Foundation for Basic Research, project no. 07-02-01464-a.

\end{document}